
\magnification=\magstep1
\hsize=15.7truecm \vsize=23.4truecm
\baselineskip=6mm
\font\bg=cmbx10 scaled 1200
\footline={\hfill\ -- \folio\ -- \hfill}
\def\prenum#1{\rightline{#1}}
\def\date#1{\rightline{#1}}
\def\title#1{\centerline{\bg#1} \vskip 5mm}
\def\author#1{\centerline{#1} \vskip 3mm}
\def\address#1{\centerline{\sl#1}}
\def\abstract#1{{\centerline{\bg Abstract}} \vskip 3mm \par #1}

\def\references{{\centerline{\bg References}} \vskip 3mm}

\def\chapter#1{\centerline{\bg#1} \vskip 5mm}
\def\section#1{\centerline{\bg#1} \vskip 5mm}
\def\endpage{\vfill \eject}

\def\footnotes#1#2{\baselineskip=5truemm \footnote{#1}{#2} \baselineskip
=6mm}
\fontdimen5\textfont2=1.2pt

\prenum{KOBE--92--03}
\date{February 1992}

\vskip 10mm

\centerline{\bg Cocycle Properties of String Theories on Orbifolds \rm
\footnotes {$^{*}$}{To appear in the proceedings of YITP Workshop on
``Recent Developments in String and Field theory" at Kyoto, Japan on
September 9-12, 1991.}}

\vskip 25mm

\author{Tsutomu HORIGUCHI,  Makoto SAKAMOTO and Masayoshi TABUSE}

\address{Department of Physics, Kobe University}
\address{Nada, Kobe 657, Japan}

\vskip 3mm

\vfill

\abstract{ We study cocycle properties of vertex operators and present an
operator representation of cocycle operators, which are attached to vertex
operators to ensure the duality of amplitudes.
 It is shown that this analysis makes it possible to obtain the general
class of consistent string theories on orbifolds.}

\endpage

\section{1. Introduction}

 Orbifold compactification [1] is believed to provide a realistic four
dimensional string model. The search for realistic orbifold models has
been continued by many authors and various models have been proposed
[2-5]. Any satisfactory orbifold models have not, however, been found yet.
So far only a very small class of orbifold models has been investigated.

In the construction of realistic four-dimensional string models, various
other approaches have been proposed [6-12].
 If string compactification can allow a geometrical interpretation,
orbifold compactification is probably the most efficient method. All $N=1$
space-time supersymmetric conformally invariant  vacua are degenerate.
The degeneracy should be resolved quantum mechanically and then a true
string vacuum will appear.
If no orbifold models were found to be realistic in spite of thorough
investigations, this might indicate that the true string vacuum is far
from all conformal invariant classical vacua and that nonperturbative
effects drastically change perturbative results [13].
If so, any conformal field  theoretical approaches would be useless to
construct a string model to describe our real world and second quantized
string field theoretical approaches [14] might be required.
In any case, more general and thorough investigations of orbifold models
would be of great importance and should be done.

  An orbifold will be obtained by dividing a torus by the action of a
discrete symmetry group $G$ of the torus.
 In ref. [15], we have clarified the general class of consistent orbifold
models:
 Any element $g$ of $G$ has been shown to be specified by
$$g=( U, v ), \eqno(1-1)$$
or more generally for asymmetric orbifolds [16]
$$g=( U_L, v_L ; U_R, v_R ), \eqno(1-2)$$
where $( U_L, U_R )$ are rotation matrices and $(v_L, v_R )$ are shift
vectors. The correct action of $g$ on a string coordinate has also been
found. In this paper, we will give some of the details of ref.[15], in
particular, a geometrical interpretation of our results and various
examples of orbifolds, which may be good illustrations of our formalism.

 In section 2, we describe the operator formalism for string theories on
orbifolds and discuss consistency conditions to determine the allowed
action of $g$ on a string coordinate.
In section 3, we investigate cocycle properties of vertex operators and
present an explicit operator representation of cocycle operators, which
are attached to vertex operators to ensure the duality of amplitudes.
 We then see that this analysis makes it possible to obtain the allowed
action of $g$ on the string coordinate and hence the general class of
consistent orbifold models.
In section 4, we discuss one loop modular invariance of partition
functions and see that this argument justifies our prescription.
In section 5, a geometrical interpretation of our results is discussed.
In section 6, we present various examples of orbifold models which may
give good illustrations of our formalism. Section 7 is devoted to
discussions.
In appendix A, various useful formulas are given and in appendix B, a part
of partition functions is explicitly evaluated.

\vskip 10mm

\section{2. Operator Formaism for Bosonic String Theories on Orbifolds}
 An orbifold [1] will be obtained by dividing a torus by the action a
suitable discrete group $G$.
Before the construction of an orbifold, we summarize the basics of strings
on a torus.
Let us start with the following action [17]\footnotes{$^*$}{$\eta^{\alpha
\beta}=diag(1,-1)$ and $\epsilon^{01}=-\epsilon^{10}=1$}
$$S=\int d\tau \int_{0}^{\pi} d\sigma {1\over 2\pi}\{\eta^{\alpha \beta}
\partial _{\alpha} X^I \partial_{\beta}X^I+\epsilon^{\alpha \beta}B^{IJ}
\partial_{\alpha}X^I \partial_{\beta}X^J\}, \eqno(2-1) $$
where $X^I(\tau,\sigma)$ $(I=1,\dots,D)$ is a string coordinate and
$B^{IJ}$ is  an antisymmetric constant background field.
 Since the second term in eq.(2-1)
is a total divergence, it does not affect the equation of motion.
 The canonical momentum conjugate to $X^I(\tau,\sigma)$, however, becomes
$$P^I(\tau , \sigma ) = {1\over \pi }({\partial}_{\tau}X^I (\tau , \sigma
) + B^{IJ} {\partial}_{\sigma}X^J (\tau , \sigma)).\eqno(2-2)$$
Thereby, the mode expansion of $X^I (\tau , \sigma)$ is given by
$$ X^I(\tau,\sigma)=x^I+(p^I-B^{IJ}w^J)\tau+w^I\sigma + {i\over 2}\sum_{n
\neq 0}{1\over n}(\alpha^I_{Ln} e^{-2in(\tau+\sigma)}+\alpha^I_{Rn}
e^{-2in(\tau-\sigma)}),\eqno(2-3)$$
where $p^I$ is the center of mass momentum and $w^I$ is the winding
number.

It is well known that the degree of freedom of the winding number must be
included in the spectrum of interacting closed strings on a torus.
 In order to construct the quantum theory, we will need to introduce a
canonical ``coordinate" $Q^I$ conjugate to $w^I$ [18]. We now assume the
following canonical commutation relations:
$$ [x^I, p^J]=i\delta^{IJ},$$
$$ [Q^I, w^J]=i\delta^{IJ}. \eqno(2-4)$$
The string coordinate $ X^{I}(\tau,\sigma)$ obeys the boundary condition
$$X^I(\tau,\sigma+\pi)=X^I(\tau,\sigma)+\pi w^I .\eqno(2-5)$$
A $D$-dimensional torus
$T^D$ may be defined by $T^D={\bf R}^D \slash \pi \Lambda $, where $
\Lambda $ is a $D$-dimensional lattice.
Since $X^I(\tau,\sigma)$ is assumed to be a string coordinate on the
torus, $w^I$ has to lie on the lattice $\Lambda$, i.e.,
$$w^I\in \Lambda .\eqno(2-6)$$
Since the wave function $\Psi(x^I)$ must be periodic, i.e., $  \Psi(x^I+
\pi w^I)=\Psi(x^I)$ for any $ w^I\in \Lambda $, the allowed momentum is
 $$p^I\in 2\Lambda^{*} ,\eqno(2-7) $$
where $\Lambda ^{*}$ is the dual lattice of $ \Lambda$.

 For later convenience, we introduce the left- and right-moving
coordinates
$$X^I(\tau,\sigma)={1\over2}(X^I_L(\tau+\sigma)+X^I_R(\tau-\sigma)),\eqno
(2-8)$$
where
$$X_L^I(\tau+\sigma)=x^I_L+2p^I_L(\tau+\sigma)+i\sum_{n\neq 0}{1\over n}
\alpha^I_{Ln}e^{-2in(\tau+\sigma)},$$
$$X_R^I(\tau-\sigma)=x^I_R+2p^I_R(\tau-\sigma)+i\sum_{n\neq 0}{1\over n}
\alpha^I_{Rn}e^{-2in(\tau-\sigma)}.\eqno(2-9)$$
The relations between $x^I,p^I,Q^I,w^I$ and $x^I_L,p^I_L,x^I_R,p^I_R$ are
given by
$$x^I_L=(1-B)^{IJ}x^J+Q^I,$$
$$x^I_R=(1+B)^{IJ}x^J-Q^I,$$
$$p^I_L={1\over2}p^I+{1\over2}(1-B)^{IJ}w^J,$$
$$p^I_R={1\over2}p^I-{1\over2}(1+B)^{IJ}w^J.\eqno(2-10)$$
Then, the commutation relations are given by
$$[x^I_L,p^J_L]=i\delta^{IJ}=[x^I_R,p^J_R],$$
$$[\alpha^I_{Lm},\alpha^J_{Ln}]=m\delta^{IJ} \delta_{m+n,0}=[
\alpha^I_{Rm},\alpha^J_{Rn}],$$
$$otherwise \ zeros.\eqno(2-11)$$
 It follows from the definition (2-10) that the left- and right-moving
momentum $(p^I_L, p^I_R)$ lies on a $(D+D)$-dimensional lorentzian even
self-dual lattice $\Gamma^{D,D}$ [17],
$$(p^I_L,p^I_R)\in \Gamma^{D,D} .\eqno(2-12)$$
 This observation is important to one loop modular invariance and also our
following discussions.

Let us introduce the complex variables $z$ and $\bar z $ defined by
$$z=e^{-2i(\tau+\sigma)},$$
$$\bar z =e^{-2i(\tau-\sigma)} .\eqno(2-13)$$
In terms of $z$ and $\bar z$, the left- and right-moving string
coordinates (2-9) can be written as
$$X^I_L(z)=x^I_L-ip^I_L lnz+i\sum_{n\neq 0}{1\over n}
\alpha^I_{Ln}z^{-n},$$
$$X^I_R(\bar z)=x^I_R-ip^I_R ln\bar z +i\sum_{n\neq 0}{1\over n}
\alpha^I_{Rn}\bar z^{-n}.\eqno(2-14)$$
 In the following analysis, the complex variable $\bar z$ will be treated
as complex conjugation of $z$ in the sense of Wick rotation.

 An orbifold is defined by specifying the action of each group element $g$
of $G$ on the left-and right-moving string coordinate $(X^I_L,X^I_R)$ $(I
=1,\dots,D)$.
 In order to determine the allowed action of $g$ on the string coordinate,
we require the following three conditions:
\item{(i)} The invariance of the energy-momentum tensors under the action
of $g$; This condition guarantees the single-valuedness of the
energy-momentum tensors on the orbifold.
\item{(ii)} The duality of amplitudes; This is one of the important
properties of string theories [19,20].
\item{(iii)} Modular invariance of partition functions; Modular invariance
plays an important role in the construction of consistent string models
[20] and conformally invariant field theories [21].
 Modular invariance may ensure the ultraviolet finiteness and the anomaly
free condition of superstring theories [20,22]. The space-time unitary
also requires modular invariance [23].

 Although the first and the third conditions (i) and (iii) have already
been investigated, no close examination has been made on the second
condition (ii) so far. As we will see later, our main results will be
obtained from the detailed analysis of the second condition(ii).

 Let us first consider the condition (i), that is, the energy-momentum
tensors have to be invariant under the action of $g$. The energy-momentum
tensors of the left- and right-movers are given by
$$T_L(z)=\lim_{w\to z}{1\over 2}P^I_L(w)P^I_L(z)-{D \over (w-z)^2},$$
$$T_R(\bar z)=\lim_{\bar w\to \bar z}{1 \over 2}P^I_R(\bar w)P^I_R(\bar
z)-{D \over (\bar w-\bar z)^2},\eqno (2-15)$$
where $P^I_L(z)$ and $P^I_R(\bar z)$ are the momentum operators of the
left- and right-movers defined by
$$\eqalign{P^I_L(z)&=i\partial_{z}X^I_L(z),\cr
P^I_R(\bar z)&=i\partial_{\bar z}X^I_R(\bar z), \quad (I=1,\dots,D).\cr}
\eqno(2-16)$$
 It follows that the energy-momentum tensors are invariant under the
action of $g$ if
$$g(P^I_L(z),P^I_R(\bar z))g^{\dagger}=(U^{IJ}_LP^J_L(z),U^{IJ}_RP^J_R(
\bar z)),\eqno(2-17)$$
where $U_L$ and $U_R$ are suitable elements of the $D$-dimensional
orthogonal group $O(D)$. Note that $U_L$ is not necessarily equal to $U_R$
and orbifolds with $U_L\neq U_R$ are called asymmetric orbifolds [16]. In
terms of $(p^I_L,\alpha^I_{Ln})$ and $(p^I_R,\alpha^I_{Rn})$, eq.(2-17)
can be rewritten as
$$\eqalign{
g(p^I_L,\alpha^I_{Ln})g^{\dagger}&=U^{IJ}_L(p^J_L, \alpha^J_{Ln}),\cr
g(p^I_R,\alpha^I_{Rn})g^{\dagger}&=U^{IJ}_R(p^J_R,\alpha^J_{Rn}).\cr}\eqno
(2-18)$$
Since $(p^I_L, p^I_R)$ lies on the lattice $\Gamma^{D,D}$, the action of
$g$ on $(p^I_L, p^I_R)$ has to be an automorphism of $\Gamma^{D,D}$, i.e.,
$$( U^{IJ}_Lp^J_L, U^{IJ}_Rp^J_R)\in \Gamma^{D,D} \quad{\rm for \ all}
\quad (p^I_L, p^I_R)\in \Gamma^{D,D}.\eqno(2-19)$$

Since the momentum operators $P^I_L(z)$ and $P^I_R(\bar z)$ do not include
$x^I_L$ and $x^I_R$, eq.(2-17) does not completely determine the action of
$g$ on $(x^I_L, x^I_R)$. In fact, the general action of $g$ on $(x^I_L,
x^I_R)$, which is compatible with the quantization conditions (2-11), may
be given by [24]
$$g(x^I_L,x^I_R)g^{\dagger}=(U^{IJ}_L(x^J_L+\pi{\partial \Phi(p_L, p_R)
\over \partial p^J_L}) ,\ U^{IJ}_R(x^J_R+\pi{\partial \Phi(p_L,p_R) \over
\partial p^J_R})),\eqno(2-20)$$
where $\Phi (p_L,p_R)$ is an arbitrary function of $p^I_L$ and $p^I_R$.
Let $g_U$ be the unitary operator which satisfies
$$g_U(X^I_L(z),X^I_R(\bar z))g_U^{\dagger}=(U^{IJ}_LX^J_L(z),U^{IJ}_RX^J_R
(\bar z)), \eqno(2-21)$$
 and
$$g_U|0> =|0>,\eqno(2-22) $$
where $ |0> $ is the vacuum of the untwisted sector. Then, the twist
operator $g$ which generates the transformations (2-18) and (2-20) will be
given by

$$g=e^{i\pi\Phi (p_L ,p_R )}g_U.\eqno(2-23)$$
 At this stage, $ \Phi $ is an arbitrary function of $p^I_L$ and $p^I_R$ .
In the next section, we will see that the second condition (ii) severely
restricts the form of the phase factor in $g$.

\vskip 10mm

\section{3. Cocycle Properties of Vertex Operators}
 In this section, we shall investigate the second condition (ii), i.e.,
the duality of amplitudes, in detail. To this end, it will be necessary to
examine cocycle properties of vertex operators and to give an explicit
operator representation of cocycle operators, which may be attached to
vertex operators.

Let us consider a vertex operator which describes the emission of a state
with  the momentum $(k^I_L, k^I_R)\in \Gamma^{D,D},$
$$V(k_L,k_R;z)=:e^{ik_L\cdot X_L(z)+ik_R\cdot X_R(\bar z)}C_{k_L,k_R}:,
\eqno(3-1)$$
where : : denotes the normal ordering and $C_{k_L,k_R}$ is the cocycle
operator which may be necessary to ensure the correct commutation
relations and the duality of amplitudes [20,25]. The product of two vertex
operators
$$V(k_L,k_R;z)V(k'_L, k'_R;z'),\eqno(3-2)$$
 is well-defined if $|z| > |z'|$. The different ordering of the two vertex
operators corresponds to the different ``time"-ordering . To obtain
scattering amplitudes, we must sum over all possible ``time"-ordering for
the emission of states. We must then establish that each contribution is
independent of the order of the vertex operators to enlarge the regions of
integrations over $z$ variables [19]. Thus the product (3-2), with respect
to $z$ and $z'$, has to be analytically continued to the region $|z'| >
|z| $ and to be identical to
$$V(k'_L, k'_R ; z')V(k_L ,k_R ;z ), \eqno(3-3)$$
for $ |z'| > |z| $.  In terms of the zero mode, the above statement can be
expressed as
$$ V_0(k_L,k_R)V_0(k'_L,k'_R)=(-1)^{k_L\cdot k'_L-k_R\cdot k'_R}V_0
(k'_L,k'_R)V_0(k_L,k_R),\eqno(3-4)$$
where
$$V_0(k_L,k_R)=e^{ik_L\cdot x_L+ik_R\cdot x_R}C_{k_L,k_R}.\eqno(3-5)$$
 This relation will follow from the following formula:
$$\eqalign{& :e^{ik_L\cdot X_L(z)+ik_R\cdot X_R(\bar z)}: :e^{ik'_L \cdot
X_L (z') + ik'_R\cdot X_R ({\bar z }')}: \cr
& =(z-z')^{k_L\cdot k'_L}(\bar z-\bar z')^{k_R\cdot k'_R} \cr
& \qquad \times :e^{ik_L\cdot X_L(z)+ik_R\cdot X_R(\bar z)+ik'_L\cdot X_L
(z')+ik'_R \cdot X_R({\bar z }')}:, \cr}  \eqno(3-6)  $$
 for $|z|>|z'|$. The factor $(-1)^{k_L\cdot k'_L-k_R\cdot k'_R}$ appearing
in eq. (3-4) is the reason for the necessity of the cocycle operator
$C_{k_L,k_R}$.

The second condition (ii) is now replaced by the statement that the
duality relation (3-4) has to be preserved under the action of $g$. To
examine this condition, we need to know an explicit operator
representation of the cocycle operator $C_{k_L,k_R}$.
For notational simplicity, we may use the following notations: $k^A \equiv
(k^I_L, k^I_R),\quad  x^A \equiv (x^I_L, x^I_R),\dots$ etc.
$(A, B, \dots$ run from 1 to $2D$ and $I, J,\dots$ run from 1 to $D$.) To
obtain an operator representation of the cocycle operator $C_k$, let us
assume [26,27]
$$C_k=e^{i\pi k^A M^{AB}{\hat p}^B} ,\eqno(3-7)$$
 where the wedge $\wedge$ may be attached to operators to distinguish
between c-numbers and q-numbers. It follows from (3-4) that the matrix
$M^{AB}$ has to satisfy
$$e^{i\pi k^A (M-M^T)^{AB}k'^B}=(-1)^{k^A \eta^{AB} k'^B}\qquad{\rm for \
all}\quad  k^A, k'^A \in  \Gamma ^{D,D},\eqno(3-8)$$
where
$$\eta^{AB}=\pmatrix{{\bf 1}&{\bf 0}\cr {\bf 0}& -{\bf 1}\cr}^{AB}.\eqno
(3-9)$$
 A solution to this equation may be given by
$$M^{AB}={\pmatrix{-{1\over 2}B^{IJ}& -{1\over 2}(1-B)^{IJ} \cr {1\over 2}
(1+B)^{IJ} &-{1\over 2}B^{IJ} }}^{AB} .\eqno(3-10)$$
To see this, first note that $M^T=-M$ and consider
$$\eqalign{2k^A M^{AB} k'^B &=-(k_L-k_R)^I ((1+B)^{IJ}k'^J_L+
(1-B)^{IJ}k'^J_R) + k^I_L k'^I_L -k^I_R k'^I_R \cr
 &=k^I_L k'^I_L-k^I_R  k'^I_R \quad{\rm mod} \ 2 ,\cr} \eqno(3-11)$$
 where we have used the fact that
$$k^I_L-k^I_R \in \Lambda ,$$
$$(1+B)^{IJ} k'^J_L + (1-B)^{IJ} k'^J_R \in 2 \Lambda^* .\eqno(3-12)$$

Although we have obtained a representation of the cocycle operator $C_k$,
its representation is not unique.
Indeed, there exist infinitely many other representations of $C_k$. In
ref. [15], it has, however, been proved that by a suitable unitary
transformation any representation of $C_k$ can reduce to eq.(3-7) with
(3-10) up to a constant phase. Thus, it will be sufficient to consider
only the representation (3-7) with (3-10) for our purpose .

 To explicitly show the dependence of the cocycle operator in the zero
mode part of the vertex operator (3-5), we may write
$$ V_0 (k ; M ) \equiv e^{ik\cdot \hat x}e^{i\pi k\cdot M \hat p}.\eqno
(3-13)$$
Under the action of $g_U$, $V_0 (k;M) $ transforms as
$$ g_UV_0 (k;M)g^{\dagger}_U =V_0 (U^T k; U^T M U),\eqno(3-14)$$
where
$$U^{AB}=\pmatrix{U^{IJ}_L & 0\cr 0 & U^{IJ} _R \cr}^{AB} .\eqno(3-15)$$
 It is easy to see that the product of $V_0 (k;M)$ and $V_0 (k' ; U^T M U
)$ satisfies
$$V_0 (k;M)V_0 (k';U^T M U )=\xi (-1)^{k\cdot \eta k'}V_0 (k';U^T M U )V_0
(k;M),\eqno(3-16)$$
where
$$\xi =e^{-i\pi k \cdot (M-U^T M U)k'}. \eqno(3-17)$$
This relation implies that the duality relation (3-4) cannot be preserved
under the action of $g_U $ unless $\xi =1$. It does not, however, mean
the violation of the duality relation under the action of $g$ because the
freedom of $\Phi (p) $ in $g$ has not been used yet. Under the action of
$g$, $V_0 (k;M) $ transforms as
$$ gV_0(k;M)g^{\dagger} =e^{i(U^T k)\cdot \hat x } e^{i\pi (U^T k)\cdot
U^T M U \hat p +i\pi \Phi(\hat p +U^T k)-i\pi \Phi (\hat p ) } .\eqno
(3-18)$$
In order for the duality relation to be preserved, we may require that
$$ gV_0 (k;M)g^{\dagger} \propto V_0 (U^T k; M ), \eqno(3-19)$$
 where the proportional constant is required to be a c-number because a
q-number phase will destroy the duality relation. Suppose that $\Phi (p)$
is expanded as
$$\eqalign{
\Phi (p) &=\phi + 2v^A \eta^{AB} p^B + {1\over 2} C^{AB} p^A p^B\cr
 &\quad + \sum _{n \geq 3} {1\over n!}C^{A_1 \dots A_n}p^{A_1}\cdots
p^{A_n}.\cr}\eqno(3-20)$$
Inserting \ eq.(3-20) \ into \ eq.(3-18) \ and \ requiring \ eq.(3-19), \
we \ may \ conclude \ that \ $C^{A_1\dots A_n}=0$ for $n \geq 3 $ and
$$k^A C^{AB} k'^B =k^A (M-U^T M U )^{AB} k'^B \quad{\rm mod}\  2 , \eqno
(3-21) $$
for $k^A,k'^A \in \Gamma^{D,D}$. There are no constraints on $\phi$ and
$v^A$. This result is nothing but the result given in ref. [15], where a
slightly different approach has been used.
It seems that there is no solution to eq.(3-21) because $C^{AB}$ is a
symmetric matrix while $M^{AB}$ is an antisymmetric one.
However, we can always find a symmetric matrix $C^{AB}$ satisfying (3-21).
To see this, let us introduce a basis $e^A_a$ $(a=1,\dots , 2D)$ of $
\Gamma^{D,D}$, i.e., $k^A=\sum_{a=1}^{2D}k^ae^A_a$ $(k^a\in {\bf Z})$ for
$k^A\in\Gamma^{D,D}$.
Then, eq. (3-21) may be rewritten as
$$C_{ab}=e^A _a (M-U^T M U )^{AB} e^B _b \quad{\rm mod}\  2 , \eqno
(3-22)$$
where $C_{ab} \equiv e^A_a C^{AB} e^B _b $.
Since the matrix $M^{AB}$ satisfies eq.(3-8), we find
$$2k^A ( M-U^T M U )^{AB} k'^B = 0 \qquad{\rm mod}\  2 , \eqno(3-23)$$
because $U^{AB} k^B,\  U^{AB}k'^B \in \Gamma^{D,D}$ and $ U^T U = {\bf
1}$.
This implies that
$$k^A (M-U^T M U )^{AB}k'^B \in {\bf Z} ,$$
or equivalently
$$ e^A_a (M - U^T M U )^{AB} e^B _b \in {\bf Z} . \eqno(3-24)$$
This guarantees the existence of a solution to eq. (3-22).

 We have observed that the duality relation can be preserved under the
action of $g$  if $\Phi (p) $ in $g$ is chosen as
$$ \Phi (p)=\phi + 2v^A \eta^{AB} p^B + {1\over 2}C^{AB} p^A p^B , \eqno
(3-25)$$
where the symmetric matrix $C^{AB} $ is defined through the relation
(3-21) or (3-22).
We will see in the next section that modular invariance requires $\phi = 0
$ and imposes some constraints on $ v^A $. The symmetric matrix $C^{AB}$
seems not to be defined uniquely in eq.(3-21) or (3-22).
Let $C'^{AB} $ be another choice satisfying eq.(3-21).
 Writing $p^A = \sum p^a e^A_a$ with $p^a \in {\bf Z}$ and defining
$C_{ab}=e_a ^A C^{AB} e_b ^B $ , we have
$$\eqalign{\qquad{1\over 2 }(C'^{AB}-C^{AB})p^A p^B & = {1 \over 2} \sum
_{a,b}(C'_{ab}-C_{ab})p^a p^b  \cr
& =  {1\over 2}\sum _{a=b}(C'_{aa}-C_{aa})(p^a)^2  + \sum _{a<b}
(C'_{ab}-C_{ab} )p^a p^b \cr
& = {1\over 2} \sum_a (C'_{aa}-C_{aa})p^a \qquad{\rm mod} \ 2,\qquad\qquad
\qquad (3-26)\cr}$$
where we have used the fact that
$C'_{ab} - C_{ab} \in 2{\bf Z} $ and $p^a \in {\bf Z }$.
 Thus, the difference between $C^{AB} $ and $C'^{AB} $ can be absorbed
into the redefinition of $v^A $ and hence the choice of $C^{AB} $ is
essentially unique.
 Therefore, we have found that any twist operator $g$ can always be
parametrized by $( U_L,v_L; U_R, v_R )$, as announced in the introduction.

\vskip 10mm

\section {4. One Loop Modular Invariance}
 In this section ,we will investigate one loop modular invariance of
partition functions. Let $Z(h,g;\tau )$ be the partition function of the
$h$-sector twisted by $g$ which is defined, in the operator formalism, by
$$Z(h,g;\tau )={\rm Tr}[g e^{i2\pi \tau (L_0 - {D\over 24})-i2\pi \bar
\tau ({\bar L}_0 -{D\over 24})}]_{h-sector} , \eqno(4-1)$$
where $L_0 (\bar L_0 )$ is the Virasoro zero mode operator of the left-
(right-) mover. The trace in eq. (4-1) is taken over the Hilbert space of
the $h$-sector. Then, the one loop partition function will be of the form,
$$Z(\tau)={1\over N} \sum _{g,h\in G \atop gh=hg}Z(h,g;\tau) , \eqno
(4-2)$$
where $N$ is the order of $G$.
In the above summation, only the elements $h$ and $g$ which commute each
other contribute to the partition function.
 This will be explained as follows: On the orbifold, each string obeys a
boundary condition such that for some element $h \in G$,
$$(X^I_L(e^{2\pi i }z) , X^I _R(e^{-2\pi i}\bar z)) = h\cdot (X^I_L (z),
X^I_R (\bar z)),\eqno (4-3)$$
up to a torus shift. A string obeying the boundary condition (4-3) is said
to belong to the $h$-sector.
If $h$ is not the unit element of $G$, such string is called a twisted
string. The total Hilbert space ${\cal H }_{total}$ consists of the direct
sum of every Hilbert space ${\cal H}_h (h\in G)$,
$${\cal H}_{total}=\bigoplus_{h\in G} {\cal H}_h . \eqno(4-4)$$
The physical Hilbert space is not the total Hilbert space itself but the
$G$-invariant subspace
of ${\cal H}_{total}$ because any physical state must be invariant under
the action of all $g \in G$. Thus, the partition function will be given by

$$Z(\tau)=\sum _{h \in G} Z(\tau)_{h-sector} , \eqno(4-5)$$
where
$$Z(\tau)_{h-sector}={\rm Tr}^{(phys)}[e^{i2\pi \tau (L_0 -{D\over 24})-i2
\pi \bar \tau(\bar L_0 -{D\over 24})}]_{h-sector}.\eqno(4-6)$$
Here, the trace should be taken over the physical Hilbert space of the
$h$-sector, which will be given by
$${\cal H}^{(phys)}_h ={\cal P} {\cal H}_h ,\eqno(4-7)$$
where ${\cal P}$ is the projection operator defined by
$${\cal P} = {1\over N}\sum_{g \in G} g .\eqno(4-8)$$
By use of the projection operator , the trace formula (4-6) may be
rewritten as
$$Z(\tau )_{h-sector}={\rm Tr}[{\cal P} e^{i2\pi \tau (L_0 -{D\over
24})-i2\pi \bar \tau (\bar L_0 -{D\over 24})}]_{h-sector},\eqno(4-9)$$
where the trace is taken over the Hilbert space ${\cal H}_h$. Let us
consider the action of $g$ on the string coordinate $(X^I _L (z) , X^I _R
(\bar z))$ in the $h$-sector.
It follows from (4-3) that $g(X^I _L(z),X^I _R ( \bar z))g^{\dagger}$
obeys the boundary condition of the $ghg^{-1}$-sector.
Let $|h>$ be any state in the $h$-sector.
The above observation may imply that $g|h>$ belongs to the
$ghg^{-1}$-sector but not the $h$-sector (unless $g$ commutes with $h$).
Therefore, in the trace formula (4-9),
$${\rm Tr}[g e^{i2\pi \tau (L_0 -{D\over 24})-i2\pi \bar \tau(\bar L_0 -{D
\over 24})}]_{h-sector},\eqno(4-10)$$
will vanish identically unless $g$ commutes with $h$.

One loop modular invariance of the partition function is satisfied
provided
  $$Z(h,g;\tau +1) = Z(h,hg;\tau ), \eqno(4-11)$$
  $$Z(h,g;-{1\over \tau}) = Z(g^{-1},h;\tau ). \eqno(4-12)$$
Let us first evaluate the partition function of the untwisted sector
twisted by $g$, $Z(1,g;\tau )$.
 It follows from the discussions of the previous section that in the
untwisted sector the twist operator $g$ would be of the form
$$g=e^{i\pi \Phi (p)}g_U  , \eqno(4-13)$$
where
$$\Phi (p) =\phi + 2v^A \eta ^{AB} p^B + {1\over 2 }p^A C^{AB} p^B .\eqno
(4-14)$$
Let $n$ be the smallest positive integer such that $g^n =1$. It means that
$$U^n ={\bf 1} , \eqno(4-15)$$
$$n\phi+\sum^{n-1}_{\ell =0}\lbrace 2v\cdot \eta U^{\ell} p + {1\over 2}p
\cdot U^{-{\ell}} C U^{\ell} p \rbrace = 0 \quad{\rm mod \ 2 \quad for\
all} \ p^A \in \Gamma ^{D,D} .\eqno(4-16)$$
The zero mode part of $Z(1,g;\tau)$ can easily be evaluated and the result
is
$$Z(1,g;\tau)_{zero\  mode} = \sum _{(k_R, k_R)\in \Gamma ^{d,\bar d}_g}
e^{i\pi \Phi (k)} e^{i\pi \tau k^2 _L -i\pi \bar \tau k^2 _R}, \eqno
(4-17)$$
where $\Gamma_g^{d,\bar d}$ is the $g$-invariant sublattice of $
\Gamma^{D,D}$, i.e.,
$$\Gamma ^{d,\bar d} _g = \lbrace (k_L,k_R) \in \Gamma ^{D,D} | (U_L k_L ,
U_R k_R) = (k_L , k_R)\rbrace . \eqno(4-18)$$
Here, $d+ \bar d$ denotes singature of the lorentzian lattice $\Gamma ^{d,
\bar d}_g$ . We now show that the following relation holds for a suitable
constant vector $v'^A $ :
$$ {1\over 2 } k^A C^{AB} k^B = 2 v'^A \eta ^{AB} k^B \quad{\rm mod} \ 2 ,
\eqno(4-19)$$
for all $k^A \in \Gamma^{d,\bar d}_g $. To show this , define
$$f(k) = {1\over 2}k^A C^{AB} k^B .\eqno(4-20)$$
Note that
$$\eqalign {k^A C^{AB} k'^B & = k^A (M-U^T M U)^{AB} k'^B \quad{\rm mod} \
2 \cr & = 0 \ {\rm mod} \ 2\quad{\rm for\  all} \ k^A , k'^A \in \Gamma
^{d,\bar d }_g ,\cr} \eqno(4-21)$$
where we have used eqs. (3-21) and (4-18) .
It follows that
$$f(k +k') = f(k) + f(k') \quad{\rm mod} \ 2 , \eqno(4-22)$$
for all $k , k' \in \Gamma ^{d,\bar d }_g $. This relation ensures the
existence of a vector $v'^A$ satisfying eq . (4-19) .
Using the relation (4-19), we can write (4-17) as
$$Z(1, g ; \tau )_{zero\  mode}  = \sum _{(k_L , k_R) \in \Gamma ^{d,\bar
d }_g} e^{i\pi \phi + i2\pi (v+v')\cdot \eta k} e^{i\pi \tau k^2_L - i\pi
\bar \tau k^2_R}. \eqno(4-23)$$
It will be useful to introduce a projection matrix ${\cal P}_U$ defined by
$${\cal P}_U = {1\over n } \sum ^{n-1} _{\ell =0} U^{\ell}. \eqno(4-24)$$
Noting that ${\cal P}_U k = k $ for all $ k \in \Gamma ^{d,\bar d }_g $
and using the Poisson resummation formula, we have
$$\eqalign{& Z(1 , g ; -{1\over \tau })_{zero\  mode } \cr
& =e^{i\pi \phi } {{(-i \tau )^{d\over 2} ({i\bar\tau } )^{\bar d \over 2}
}\over V_{\Gamma ^{d,\bar d}_g}} \sum _{(q_L , q_R) \in \Gamma ^{d,{\bar
d}^*}_g - (v^* + v'^*)} e^{i\pi \tau q^2_L - i\pi \bar \tau q^2_R },\cr}
\eqno(4-25)$$
where $v^* + v'^* \equiv {\cal P}_U (v + v')$,
 $V_{\Gamma}$ is the unit volume of the lattice $\Gamma $ and $\Gamma
^{d,{\bar d}^* } _g $ is the dual lattice of $\Gamma ^{d,{\bar d} }_g $.
It follows from eq.(4-25) that we can easily extract information about the
zero mode of the $g^{-1}$-sector because $Z(1,g;\tau )$
should be related to $Z(g^{-1} , 1;\tau )$ thorough the modular
transformation, i.e.,
$$Z(g^{-1} , 1 ; \tau )= Z(1 , g ; -{1 \over \tau }) . \eqno (4-26)$$
It turns out that the degeneracy of the ground state in the
$g^{-1}$-sector may be given by [16]
$${\sqrt{det'(1-U)}\over V_{\Gamma ^{d,\bar d}_g } } , \eqno(4-27)$$
where the determinant should be taken over the nonzero eigenvalues of
$1-U$ and the factor $\sqrt{det'(1-U)}$ will come from the oscillators.
The eigenvalues of the momentum $(q_L, q_R)$ in the $g^{-1}$-sector may be
given by
$$(q_L , q_R ) \in \Gamma^{d,\bar d*}_g - v^* - v'^* . \eqno(4-28)$$

It should be noted that the momentum eigenvalues in the $g^{-1}$-sector
are not given by $\Gamma_g^{d,\bar d *} - v^* $, which might naively be
expected [16].
The origin of the extra contribution $-v'^*$ is the third term in eq.
(4-14), which has been introduced to ensure the duality relation of vertex
operators.  As we will see later, this extra contribution to the momentum
eigenvalues becomes important to ensure the left-right level matching
condition.

Information about the zero mode given above is sufficient to obtain $Z
(g^{-1} , 1 ; \tau )$ because the oscillator part of $Z(g^{-1} , 1 ; \tau
)$ can unambiguously be calculated .
In appendix B , we will prove that the relation (4-26) puts a constraint
on $\phi $ in eq. (4-14), i.e.,
$$\phi =0 . \eqno (4-29)$$
This is desirable because otherwise the vacuum in the untwisted sector
would not be invariant under the action of $g$ and hence would be removed
from the physical Hilbert space.
In the point of view of the conformal field theory the vacuum in the
untwisted sector will correspond to the identity operator,
which should be included in the operator algebra.

A necessary condition for modular invariance is the left-right level
matching condition [16,28]
$$Z(g^{-1},h;\tau+n)=Z(g^{-1},h;\tau).\eqno(4-30)$$
It follows from eq.(4-1) that the level matching condition is satisfied
only if
$$2n(L_0-\bar L_0)=0 \quad {\rm mod} \ 2,\eqno(4-31)$$
where $L_0$ $(\bar L_0)$ is the Virasoro zero mode operator of the left-
(right-) mover in the $g^{-1}$-sector. Since any contribution to $L_0$ and
$\bar L_0$ from the oscillators is a fraction of $n$, the level matching
condition can be written as
$$2n(\varepsilon_{g^{-1}}-\bar\varepsilon_{g^{-1}}+{1\over2}q_L^2-{1
\over2}q_R^2)=0 \quad {\rm mod} \ 2, \qquad {\rm for \ all} \ (q_L,q_R)\in
\Gamma^{d,\bar d^*}_g-v^*-{v'}^*,\eqno(4-32)$$
where $(\varepsilon_{g^{-1}},\bar\varepsilon_{g^{-1}})$ is the conformal
dimension (or the zero point energy) of the ground state in the
$g^{-1}$-sector and is explicitly given by [1]
$$\varepsilon_{g^{-1}}={1\over4}\sum_{a=1}^D\rho_a(1-\rho_a),$$
$$\bar\varepsilon_{g^{-1}}={1\over4}\sum_{a=1}^D\bar\rho_a(1-\bar\rho_a).
\eqno(4-33)$$
Here, $exp(i2\pi\rho_a)$ and $exp(i2\pi\bar\rho_a)$ $(a=1,\cdots,D)$ are
the eigenvalues of $U_L$ and $U_R$ with $0\le\rho_a,\bar\rho_a<1$,
respectively. The condition (4-32) can further be shown to reduce to
$$2n(\varepsilon_{g^{-1}}-\bar\varepsilon_{g^{-1}}+{1\over2}(v^*_L
+{v'}_L^*)^2-{1\over2}(v^*_R+{v'}_R^*)^2)=0 \quad {\rm mod} \ 2.\eqno
(4-34)$$
To see this, we first note that ${\Gamma^{d,\bar d}_g}^*$ can be expressed
as [16]
$$\eqalign{ {\Gamma^{d,\bar d}_g}^*&={\cal P}_U\Gamma^{D,D}\cr
                                   &=\{q^A={\cal P}_Uk^A, \ k^A\in
\Gamma^{D,D}\}.}\eqno(4-35)$$
This follows from the property that $\Gamma^{D,D}$ is self-dual. From eq.
(4-35), any momentum $q\in{\Gamma^{d,\bar d}_g}^*-v^*-{v'}^*$ can be
parametrized as
$$q^A={\cal P}_U(k-v-v')^A \quad {\rm for \ some }\ k^A\in\Gamma^{D,D}.
\eqno(4-36)$$
Then, we have
$$\eqalign{\quad n(q_L^2-q_R^2)&=nq\cdot\eta q\cr
                         &=nk\cdot\eta{\cal P}_Uk-2n(v+v')\cdot\eta{\cal
P}_Uk+n(v^*+{v'}^*)\cdot\eta(v^*+{v'}^*),\quad (4-37)\cr}$$
where we have used the relations
$${\cal P}_U\eta=\eta{\cal P}_U,$$
$${\cal P}_U^2={\cal P}_U,$$
$${\cal P}_U^T={\cal P}_U.\eqno(4-38)$$
Since $\Gamma^{D,D}$ is an even integral lattice and $U$ is an orthogonal
matrix satisfying $U^n={\bf 1}$, the first term in the right handed side
of eq.(4-37) is easily shown to reduce to
$$ nk\cdot\eta{\cal P}_Uk=\cases{k\cdot\eta U^{n\over2}k \quad & mod 2  if
 $n=\rm even$,\cr
0 \quad & mod  2  if $n=\rm odd$. \cr} \eqno(4-39)$$
Using the relation (4-19) and nothing that $n{\cal P}_Uk\in\Gamma^{d,\bar
d}_g$, we can rewrite the second term in the right hand side of eq.(4-37)
as
$$-2n(v+v')\cdot\eta{\cal P}_Uk=-2nv\cdot\eta{\cal P}_Uk-{1\over2}k\cdot(
\sum_{\ell=0}^{n-1}U^{-\ell})C(\sum_{m=0}^{n-1}U^m)k \quad{\rm mod} \ 2.
\eqno(4-40)$$
Replacing $p$ by $p+p'$ in eq.(4-16) with eq.(4-29) and then using (4-16)
again, we have
$$p\cdot\sum_{\ell=0}^{n-1}U^{-\ell}CU^\ell p'=0 \quad {\rm mod} \ 2,
\eqno(4-41)$$
for all $p,p'\in\Gamma^{D,D}$. For $n$ odd, it is not difficult to show
that
$$-2n(v+v')\cdot\eta{\cal P}_Uk=0 \quad {\rm mod} \ 2.\eqno(4-42)$$
To derive eq.(4-42), we will use eqs.(4-16),(4-29),(4-40) and (4-41). For
$n$ even, we will find
$$-2n(v+v')\eta{\cal P}_Uk=-k\cdot(\sum_{\ell=0}^{{n\over2}-1}U^{-\ell}CU^
\ell)U^{n\over2}k \quad {\rm mod} \ 2.\eqno(4-43)$$
Remembering eqs.(3-8) and (3-21), we can finally find that for $n$ even
$$-2n(v+v')\eta{\cal P}_Uk=-k\cdot\eta U^{n\over2}k \quad {\rm mod} \ 2.
\eqno(4-44)$$
Combining the results (4-39), (4-42) and (4-44), we have
$$nk\cdot\eta{\cal P}_Uk-2n(v+v')\cdot\eta{\cal P}_Uk=0 \quad {\rm mod} \
2.\eqno(4-45)$$
This completes the proof of (4-34).

We have shown that the left-right level matching condition (4-30) reduces
to the condition (4-34), which may put a constraint on the shift vector $v
=(v_L,v_R)$. It should be noticed that the level matching condition (4-34)
is not always satisfied for asymmetric orbifold models but trivially
satisfied for symmetric ones because $\varepsilon_{g^{-1}}=\bar
\varepsilon_{g^{-1}}$ and $(v_L^*+{v'}_L^*)^2=(v_R^*+{v'}_R^*)^2$ for
symmetric orbifold models. For the case of $C^{AB}=0$ in eq.(4-14), it has
been proved, in refs. [16,28], that the level matching condition is a
necessary and also sufficient condition for one loop modular invariance.
Even in the case of the general twist (4-13) with the phase (4-14), the
sufficiency can probably be shown by arguments similar to refs. [16,28].
It should be emphasized that the third term in eq.(4-14) plays an
important role in the level matching condition because the relation (4-45)
might not hold in general if we put $v'$ to be zero, that is, $C^{AB}$ to
be zero by hand. In section 6, we will see examples of orbifold models
that the introduction of the third term in eq.(4-14) makes partition
functions modular invariant.

Before closing this section, we shall make a comment on modular
invariance of correlation functions. Our analysis implies that one loop
modular invariance of partition functions does not in general ensure one
loop modular invariance of correlation functions because a wrong choice of
a twist operator $g$ could destroy the duality relation of vertex
operators even though the partition function is modular invariant. Such an
example will be found in section 6.

\vskip 10mm

\section{5. A Geometrical Interpretation}

We have found that the string coordinate $X^A=(X^I_L,X^I_R)$ in the
untwisted sector transforms under the action of $g$ as
$$ gX^Ag^\dagger = U^{AB}(X^B+2\pi\eta^{BC}v^C+\pi C^{BC}p^C). \eqno
(5-1)$$
It seems that the third term in the right hand side of eq.(5-1) has no
clear geometrical meaning. Although the momentum and vertex operators
definitely transform under the action of $g$, why does not the string
coordinate $(X^I_L,X^I_R)$ transform definitely ? The reason may be that
in the point of view of the conformal field theory the string coordinate
is not a primary field and it is not a well-defined variable on a torus.
Thus, there is probably no reason why the string coordinate itself should
definitely transform under the action of $g$. On the other hand, since the
momentum and vertex operators are primary fields and are well-defined on a
torus, they should definitely transform under the action of $g$. In fact,
they transform as
$$ g(P^I_L(z),P^I_R(\bar z))g^\dagger = (U^{IJ}_LP^J_L(z),U^{IJ}_RP^J_R(
\bar z)),$$
$$ gV(k_L,k_R;z)g^\dagger = e^{i2\pi v\cdot\eta U^Tk+i{\pi\over 2}k\cdot
UCU^Tk}V(U_L^Tk_L,U_R^Tk_R;z). \eqno(5-2)$$

As mentioned above, not the string coordinate but the momentum and vertex
operators are relevant operators on tori or orbifolds. Since $P^I_L(z)$
and $P^I_R(\bar z)$ do not include the ``center of mass coordinate'' $
(x^I_L,x^I_R)$, it appears only in the vertex operators. The cocycle
operator has been shown to be represented as
$$ C_k=e^{i\pi k^AM^{AB}p^B}. \eqno(5-3)$$
Therefore, we observe that the ``center of mass coordinate'' $x^A=
(x^I_L,x^I_R)$ always appears as the following combination:
$$ x^A+\pi M^{AB}p^B . \eqno(5-4)$$
This observation strongly suggests that the combination is a more
fundamental variable than $x^A$ itself. To see this, let us introduce the
variable $x'^A$, which is slightly different from the variable (5-4),
$$ x'^A \equiv x^A + \pi M'^{AB}p^B, \eqno(5-5) $$
where
$$ M'^{AB} = M^{AB} + {1\over 2} \eta^{AB}. \eqno(5-6)$$
Note that $x'^A$ is related to the variable (5-4) by the following unitary
transformation:
$$ {\cal U}(x^A+\pi M^{AB}p^B){\cal U}^\dagger = x'^A, \eqno(5-7)$$
where
$$ {\cal U}=e^{-i{\pi\over4}p^A\eta^{AB}p^B}. \eqno(5-8)$$
Hence, we will discuss a geometrical meaning of $x'^A$ instead of the
variable (5-4) in the following. We first note that although $x^A$ does
not transform definitely under the action of $g$, $x'^A$ does:
$$ gx'^Ag^\dagger \sim U^{AB}(x'^B+2\pi\eta^{BC}v^C), \eqno(5-9)$$
where $\sim$ means that the right hand side is identical to the left hand
side up to a torus shift.\footnotes{$^{*}$}{In fact, $x'^A$ transforms as
$$ gx'^Ag^\dagger = U^{AB}(x'^B+2\pi\eta^{BC}v^C)+\pi U^{AB}(U^TMU-M
+C)^{BC}p^C.$$
The last term of the right hand side is nothing but a torus shift because
$$ k^AU^{AB}(U^TMU-M+C)^{BC}p^C=0 \quad {\rm mod} \ 2 \quad {\rm for \ any
\ } k^A,p^A\in \Gamma^{D,D},$$
where we have used eq.(3-21).} In terms of the left- and right-moving
coordinates, eq.(5-5) is written as
$$(x'^I_L,x'^I_R)=(x^I_L+{\pi\over 2}(1-B)^{IJ}(p^J_L-p^J_R),x^I_R+{\pi
\over 2}(1+B)^{IJ}(p^J_L-p^J_R)).\eqno(5-10)$$
We may further rewrite the variables $x^I_L$, $p^I_L$, $x^I_R$ and $p^I_R$
into $x^I$, $p^I$, $Q^I$ and $w^I$, which will geometrically be more
fundamental than $x^I_L$, $p^I_L$, $x^I_R$ and $p^I_R$. Then, we have
$$ x'^I=x^I+{\pi\over 2}w^I, \eqno(5-11)$$
$$ Q'^I=Q^I, \eqno(5-12)$$
where $x'^I$ and $Q'^I$ are related to $x'^I_L$ and $x'^I_R$ through the
same relations as eq.(2-10). The question is now what geometrical meaning
$x'^I$ has.

Before we answer the question, it may be instructive to make a comment on
the center of mass coordinate of a string on a torus, which has a clear
geometrical meaning if the string has no winding number. The ``center of
mass coordinate'' is, however, ill-defined geometrically if the string
winds around the torus. Thus, $x^I$ can be interpreted as the center of
mass coordinate in the absence of the winding number but it will lose its
geometrical meaning in the presence of the winding number. However, it may
still be a useful notion on the covering space of the torus. It turns out
that on the covering space of the torus the ``center of mass coordinate''
of the string may be locate at [27]
$$ x'^I=x^I+{\pi\over 2}w^I. \eqno(5-13)$$
To see this, consider the string coordinate $X^I(\tau ,\sigma )$ at $\tau
= 0$ given in eq.(2-3) and integrate it over the $\sigma$-variable. Then
we have
$$ \int_0^\pi {d\sigma\over\pi}X^I(0,\sigma)=x^I+{\pi\over 2} w^I.\eqno
(5-14)$$

The above observation may suggest that the reason why cocycle operators
appear in vertex operators is related to the fact that there is no good
variable of the ``center of mass coordinate'' of a string on a torus and
also suggest that the variable $x'^I$ defined in eq.(5-11) is more
fundamental than $x^I$ on a torus as well as on an orbifold because $x'^A$
but not $x^A$ definitely transforms under the action of $g$.

\vskip 10mm

\section{6. Example of Orbifolds}

In this section, we shall investigate a symmetric ${\bf Z}_2$-orbifold, a
nonabelian $S_3$-orbifold and an asymmetric ${\bf Z}_3$-orbifold, in
detail, which will give good illustrations of our formalism.

Let us introduce the root lattice $\Lambda_R$ and the weight lattice of
$SU(3)$ as
$$\Lambda_R=\{p^I=\sum_{i=1}^2 n^i\alpha^I_i, n^i \in {\bf Z}\},$$
$$\Lambda_W=\{p^I=\sum_{i=1}^2 m_i\mu^{iI}, m_i \in {\bf Z}\},\eqno(6-1)$$
where $\alpha_i$ and $\mu^i$ $(i=1,2)$ are a simple root and a fundamental
weight satisfying $\alpha_i\cdot\mu^j=\delta_i^j$. We will take $\alpha_i$
and $\mu^i$ to be
$$\alpha_1=({1\over\sqrt 2},\sqrt{3\over2}),$$
$$\alpha_2=({1\over\sqrt 2},-\sqrt{3\over2}),$$
$$\mu^1=({1\over\sqrt 2},\sqrt{1\over6}),$$
$$\mu^2=({1\over\sqrt 2},-\sqrt{1\over6}).\eqno(6-2)$$
Let $p^I$ and $w^I$ be the center of mass momentum and the winding number,
respectively. They are assumed to lie on the following lattices:
$$p^I\in 2\Lambda_W,$$
$$w^I\in \Lambda_R.\eqno(6-3)$$
The left- and right-moving momentum $(p^I_L,p^I_R)$ is defined by eq.
(2-10), i.e.,
$$p^I_L={1\over2}p^I+{1\over2}(1-B)^{IJ}w^{J},$$
$$p^I_R={1\over2}p^I-{1\over2}(1+B)^{IJ}w^{J}.\eqno(6-4)$$
The antisymmetric constant matrix $B^{IJ}$ is chosen as
$$B^{IJ}=\pmatrix{0 & -{1\over\sqrt{3}} \cr
                  {1\over\sqrt{3}} & 0 \cr}.\eqno(6-5)$$
Then, it turns out that $(p^I_L,p^I_R)$ lies on the following $2
+2$-dimensional lorentzian even self-dual lattice:
$$\Gamma^{2,2}=\{(p^I_L,p^I_R)|p^I_L,p^I_R\in \Lambda_W, p^I_L-p^I_R\in
\Lambda_R\}.\eqno(6-6)$$

\vskip 10mm

\noindent{\bf 6-1. A Symmetric ${\bf Z}_2$-Orbifold}

We shall first consider a symmetric $SU(3)/{\bf Z}_2$-orbifold whose ${
\bf Z}_2$-transformation is defined by
$$g_U(X^I_L,X^I_R)g_U^\dagger=(U^{IJ}_LX^J_L,U^{IJ}_RX^J_R), \quad (I
=1,2),\eqno(6-7)$$
where
$$U^{IJ}_L=U^{IJ}_R=\pmatrix{-1 & 0 \cr
                             0  & 1 \cr }. \eqno(6-8)$$
This is an automorphism of $\Gamma^{2,2,}$, as it should be. According to
our prescription, the ${\bf Z}_2$-twist operator $g$ will be given by
$$g=e^{i{\pi\over2}p^AC^{AB}p^{B}}g_U,\eqno(6-9)$$
where $p^A=(p^I_L,p^I_R)$ and the symmetric matrix $C^{AB}$ is defined
through the relation
$$p^A(M-U^TMU)^{AB}p'^B=p^AC^{AB}p'^B \quad {\rm mod} \ 2,\eqno(6-10)$$
for $p^A, p'^A \in \Gamma^{2,2}$. Here, we have taken a shift vector to
zero for simplicity and $M^{AB}$, $U^{AB}$ are defined by
$$M^{AB}=\pmatrix{-{1\over2}B^{IJ} & -{1\over2}(1-B)^{IJ} \cr
                  {1\over2}(1+B)^{IJ} & -{1\over2}B^{IJ} \cr}^{AB}, $$
$$U^{AB}=\pmatrix{U^{IJ}_L & 0 \cr
                  0  &  U^{IJ}_R  \cr}^{AB}.\eqno(6-11)$$
For symmetric orbifolds $(U_L=U_R)$, the defining relation (6-10) of
$C^{AB}$ may be replaced by
$${1\over2}(p_L-p_R)^I(B-U^T_LBU_L)^{IJ}(p'_L-p'_R)^J=(p_L-p_R)^IC^{IJ}
(p'_L-p'_R)^J \quad {\rm mod} \ 2, \eqno(6-12)$$
where $C^{AB}$ has been assumed to be of the form
$$C^{AB}=\pmatrix{-C^{IJ} & C^{IJ} \cr
                  C^{IJ} & -C^{IJ} \cr}^{AB}.\eqno(6-13)$$
Then, eq.(6-9) can be written as
$$g=e^{-i{\pi\over2}(p_L-p_R)^IC^{IJ}(p_L-p_R)^J}g_U.\eqno(6-14)$$
Since $p^I_L-p^I_R\in\Lambda_R$, the equation (6-12) may be rewritten as
$${1\over2}\alpha^I_i(B-U^T_LBU_L)^{IJ}\alpha^J_j=\alpha^I_iC^{IJ}
\alpha^J_j \quad {\rm mod} \ 2.\eqno(6-15)$$
The left hand side of eq.(6-15) is found to be
$${1\over2}\alpha^I_i(B-U^T_LBU_L)^{IJ}\alpha^J_j=\pmatrix{0 & 1 \cr
                                                          -1 & 0
\cr}_{ij},\eqno(6-16)$$
and hence $C^{IJ}$ cannot be chosen to be zero. We may choose $C^{IJ}$ as
$$\alpha^I_iC^{IJ}\alpha^J_j=\pmatrix{0 & 1 \cr
                                      1 & 0 \cr}_{ij},$$
or
$$C^{IJ}=\pmatrix{1 & 0 \cr
                  0 & -{1\over3} \cr}^{IJ}.\eqno(6-17)$$
This choice turns out to be consistent with $g^2=1$.

Let us consider the following momentum and vertex operators of the
left-mover:
$$P^I_L(z)=i\partial_zX^I_L(z),$$
$$V_L(\alpha ; z)=: e^{i\alpha\cdot X_L(z)}C_\alpha:,\eqno(6-18)$$
where $\alpha$ is a root vector of $SU(3)$ and $C_\alpha$ denotes a
cocycle operator. These operators form level one Ka{\v c}-Moody algebra $
\widehat{su(3)}_{k=1}$ [25]. Under the action of $g$, they transform as
$$gP^I_L(z)g^\dagger=U^{IJ}_LP^J_L(z),$$
$$gV_L(\pm\alpha_1;z)g^\dagger=V_L(\mp\alpha_2;z),$$
$$gV_L(\pm\alpha_2;z)g^\dagger=V_L(\mp\alpha_1;z),$$
$$gV_L(\pm(\alpha_1+\alpha_2);z)g^\dagger=-V_L(\mp(\alpha_1+\alpha_2);z).
\eqno(6-19)$$
Thus, the ${\bf Z}_2$-invariant physical generators may be given by
$$\eqalign{
J_3(z)&={\sqrt{3}\over2} \{P^2_L(z)-{i\over\sqrt{6}}(V_L(\alpha_1+
\alpha_2;z)-V_L(-\alpha_1-\alpha_2;z))\},\cr
J_\pm(z)&={1\over\sqrt2}(V_L(\pm\alpha_1;z)+V_L(\mp\alpha_2;z)),\cr
J(z)&={1\over2} \{P^2_L(z)+i\sqrt{3\over2}(V_L(\alpha_1+\alpha_2;z)-V_L(-
\alpha_1-\alpha_2;z))\},\cr}\eqno(6-20)$$
which are found to form Ka{\v c}-Moody algebra $\widehat{su(2)}_{k=1}
\oplus\widehat{u(1)}$.

We now examine one loop modular invariance of the partition function
which will be given by
$$Z(\tau)={1\over2}\sum_{\ell,m=0}^1Z(g^\ell,g^m;\tau),\eqno(6-21)$$
where
$$Z(g^\ell,g^m;\tau)={\rm Tr}[g^me^{i2\pi\tau(L_0-{D\over24})-i2\pi{\bar
\tau}(\bar L_0-{D\over24})}]_{g^\ell{\rm -sector}}.\eqno(6-22)$$
The partition functions of the untwisted sector can easily be evaluated
and the result is
$$Z(1,1;\tau)={1\over|\eta(\tau)|^4}\sum_{(k_L,k_R)\in\Gamma^{2,2}}e^{i\pi
\tau k^2_L-i\pi{\bar\tau}k^2_R},\eqno(6-23)$$
$$Z(1,g;\tau)={|\vartheta_3(0|\tau)\vartheta_4(0|\tau)|\over|\eta(
\tau)|^4}\sum_{(k_L,k_R)\in\Gamma^{1,1}_g}e^{i2\pi(v'_Lk_L-v'_Rk_R)}e^{i
\pi\tau k^2_L-i\pi{\bar\tau}k^2_R},\eqno(6-24)$$
where
$$v'_L=v'_R={1\over2\sqrt6},$$
$$\Gamma^{1,1}_g=\{(k_L,k_R)=(\sqrt6n+\lambda,\sqrt6n'+\lambda),\lambda=0,
\pm\sqrt{2\over3},n,n'\in{\bf Z}\}.\eqno(6-25)$$
Here, $\eta(\tau)$ is the Dedekind $\eta$-function and $\vartheta_a(\nu|
\tau)$ $(a=1,\cdots,4)$ is the Jacobi theta function. Their definition and
properties will be found in appendix A. The shift vector $(v'_L,v'_R)$ has
been introduced through the relation (4-19).

It follows from the arguments given in section 4 that the degeneracy of
the ground state in the $g$-sector is
$${\sqrt{det'(1-U)}\over V_{\Gamma^{1,1}_g}}=1,\eqno(6-26)$$
and that the momentum eigenvalues will be given by
$$(q_L,q_R)\in{\Gamma^{1,1}_g}^*-(v'_L,v'_R),\eqno(6-27)$$
where
$${\Gamma^{1,1}_g}^*=\{(q_L,q_R)=(\sqrt{3\over2}n+\lambda,\sqrt{3\over2}n'
+\lambda),\lambda=0,\pm{1\over\sqrt6},n,n'\in{\bf Z}\}.\eqno(6-28)$$
This information is enough to obtain $Z(g,1;\tau)$ and $Z(g,g;\tau)$,
$$Z(g,1;\tau)={|\vartheta_3(0|\tau)\vartheta_2(0|\tau)|\over2|\eta(
\tau)|^4}\sum_{(q_L,q_R)\in{\Gamma^{1,1}_g}^*-(v'_L,v'_R)}e^{i\pi\tau
q^2_L-i\pi{\bar\tau}q^2_R},$$
$$Z(g,g;\tau)={|\vartheta_4(0|\tau)\vartheta_2(0|\tau)|\over2|\eta(
\tau)|^4}\sum_{(q_L,q_R)\in{\Gamma^{1,1}_g}^*-(v'_L,v'_R)}e^{i\pi
(q_L^2-q_R^2)}e^{i\pi\tau q^2_L-i\pi{\bar\tau}q^2_R}.\eqno(6-29)$$
It is easily verified from the formulas in appendix A that $Z(g^\ell,g^m;
\tau)$ satisfies the following desired relations:
$$Z(g^\ell,g^m;\tau+1)=Z(g^\ell,g^{m+\ell};\tau),$$
$$Z(g^\ell,g^m;-{1\over\tau})=Z(g^{-m},g^\ell;\tau),\eqno(6-30)$$
and hence the partition function (6-21) is modular invariant. It should be
emphasized that the existence of the shift vector $(v'_L,v'_R)$ makes the
partition function modular invariant: The level matching condition
$$Z(g,1;\tau+2)=Z(g,1;\tau),\eqno(6-31)$$
is satisfied because for all $(q_L,q_R)\in{\Gamma^{1,1}_g}^*-(v'_L,v'_R)$,
$$4({1\over2}q_L^2-{1\over2}q_R^2)=0 \quad {\rm mod } \ 2.\eqno(6-32)$$
If we put the shift vector $(v'_L,v'_R)$ or $C^{IJ}$ in $g$ to be zero by
hand, the level matching condition might, however, be destroyed because
eq.(6-32) dose not hold.

\vskip 10mm
\noindent{\bf 6-2. A Nonabelian $S_3$-Orbifold}

The next example is a nonabelian $SU(3)/S_3$-orbifold, where $S_3$ is the
symmetric group of order three. The symmetric group $S_3$ consists of six
elements $U_i$ $(i=0,\cdots,5)$,
$$U_0=\pmatrix{1 & 0 \cr
               0 & 1 \cr},$$
$$U_1=\pmatrix{-1 & 0 \cr
               0 & 1 \cr}\equiv U,$$
$$U_2=\pmatrix{-{1\over2} & -{\sqrt3\over2} \cr
               {\sqrt3\over2} & -{1\over2} \cr}\equiv V,$$
$$U_3=V^2,$$
$$U_4=VU,$$
$$U_5=UV.\eqno(6-33)$$
The matrices $U_1$, $U_4$ and $U_5$ correspond to the Weyl reflections of
$SU(3)$ with respect to the root vectors $\alpha_1+\alpha_2$, $\alpha_1$
and $\alpha_2$, respectively, and the matrices $U_2$ and $U_3$ correspond
to the rotation by ${2\pi\over3}$ and ${4\pi\over3}$, respectively. The
action of ${g_U}_i$ $(i=0,\cdots,5)$ on the string coordinate is defined
by
$${g_U}_i(X^I_L,X^I_R){{g_U}_i}^\dagger=U_i^{IJ}(X^J_L,X^J_R), \ (i=0,
\cdots,5).\eqno(6-34)$$
Each element of $S_3$ is an automorphism of $\Lambda_R$ and $\Lambda_W$
and hence $\Gamma^{2,2}$. The matrices $U$ and $V$ satisfy
$$U^2=V^3={\bf 1},$$
$$VUV=U.\eqno(6-35)$$
According to our prescription, we may write the twist operators $g_1$ and
$g_2$ which correspond to $U_1$ and $U_2$, respectively, as
$$g_1=e^{i{\pi\over2}(p_L-p_R)^IC_1^{IJ}(p_L-p_R)^J}{g_U}_1,$$
$$g_2=e^{i{\pi\over2}(p_L-p_R)^IC_2^{IJ}(p_L-p_R)^J}{g_U}_2,\eqno(6-36)$$
where the symmetric matrices $C_1$ and $C_2$ are defined by
$$\alpha^I_iC^{IJ}_1\alpha^J_j={1\over2}\alpha^I_i(B-U^T_1BU_1)^{IJ}
\alpha^J_j \quad {\rm mod} \ 2,$$
$$\alpha^I_iC^{IJ}_2\alpha^J_j={1\over2}\alpha^I_i(B-U^T_2BU_2)^{IJ}
\alpha^J_j \quad {\rm mod} \ 2,\eqno(6-37)$$
and we have put shift vectors to zero. Other twist operators will be
defined by $g_0=1$, $g_3=(g_2)^2$, $g_4=g_2g_1$ and $g_5=g_1g_2$. Explicit
calculations show that
$$\eqalign{
{1\over2}\alpha^I_i(B-U^T_1BU_1)^{IJ}\alpha^J_j&=\pmatrix{0 & 1 \cr
                                                          -1 & 0
\cr}_{ij},\cr
{1\over2}\alpha^I_i(B-U^T_2BU_2)^{IJ}\alpha^J_j&=\pmatrix{0 & 0 \cr
                                                           0 & 0
\cr}_{ij}.\cr}\eqno(6-38)$$
In order for $g_i$'s to form the symmetric group $S_3$, we may choose
\footnotes{$^{*}$}{If we choose $C_1$ and $C_2$, in general, as
$$\alpha^I_iC^{IJ}_1\alpha^J_j=\pmatrix{2m_1  & 1+2m_3 \cr
                                     1+2m_3 & 2m_2 \cr}, \quad m_i\in {\bf
Z},$$
$$\alpha^I_iC^{IJ}_2\alpha^J_j=\pmatrix{2n_1 & 2n_3 \cr
                                      2n_3 & 2n_2 \cr}, \quad n_i\in{\bf
Z},$$
with $m_1+m_2\in 2{\bf Z}$ and $m_2+n_2\in 2{\bf Z}+1$, $g_i$ $(i=0,
\cdots,5)$ forms the symmetric group $S_3$ and any choice will lead to the
same result.}
$$\alpha^I_iC^{IJ}_1\alpha^J_j=\pmatrix{0  & 1 \cr
                                      1  & 0  \cr}_{ij},$$
$$\alpha^I_iC^{IJ}_2\alpha^J_j=\pmatrix{0 & 0 \cr
                                      0 & 2 \cr}_{ij}.\eqno(6-39)$$

Since the symmetric group $S_3$ is nonabelian, the one loop partition
function will be of the form,
$$Z(\tau)={1\over6}\sum_{g_i,g_j\in S_3\atop g_ig_j=g_jg_i}Z(g_i,g_j;
\tau).\eqno(6-40)$$
It is not difficult to show that the following combinations of $Z(g_i,g_j;
\tau)$'s are modular invariant:
$$\eqalign{ &1) \ Z(1,1;\tau), \cr
            &2) \ Z(1,g_1;\tau)+Z(g_1,1;\tau)+Z(g_1,g_1;\tau), \cr
            &3) \ Z(1,g_4;\tau)+Z(g_4,1;\tau)+Z(g_4,g_4;\tau), \cr
            &4) \ Z(1,g_5;\tau)+Z(g_5,1;\tau)+Z(g_5,g_5;\tau), \cr
            &5) \ \sum_{j=2,3}Z(1,g_j;\tau)+\sum_{j=0,2,3}(Z(g_2,g_j;\tau)
+Z(g_3,g_j;\tau)). \cr}\eqno(6-41)$$
Therefore, the partition function (6-40) is also modular invariant.
\footnotes{$^{*}$}{In ref.[29], the authors have not succeeded in
obtaining a modular invariant partition function of the nonabelian
$S_3$-orbifold model because of a wrong choice of the twist operators.}
Note that 1)+2) is nothing but the partition function of the ${\bf
Z}_2$-orbifold discussed in the previous example {\bf 6-1} up to an
overall normalization. The combination 1)+3) ( 1)+4) ) is identical to 1)
+2) and corresponds to the partition function of the ${\bf Z}_2$-orbifold
associated with the Weyl reflection with respect to $\alpha_1$ ($
\alpha_2$). The combination 1)+5) corresponds to the partition function of
the ${\bf Z}_3$-orbifold whose ${\bf Z}_3$-transformation is generated by
$g_2$.

\vskip 10mm
\noindent{\bf 6-3. An Asymmetric ${\bf Z}_3$-Orbifold}

The final example is an asymmetric $SU(3)/{\bf Z}_3$-orbifold whose ${\bf
Z}_3$-transformation is defined by
$$g_U(X^I_L,X^I_R)g_U^\dagger=(U^{IJ}_LX^J_L,U^{IJ}_RX^J_R),\eqno(6-42)$$
where
$$U^{IJ}_L=\pmatrix{-{1\over2} & -{\sqrt3\over2} \cr
                     {\sqrt3\over2} & -{1\over2} \cr},$$
$$U^{IJ}_R=\pmatrix{1 & 0 \cr
                    0 & 1 \cr}.\eqno(6-43)$$
This is also an automorphism of $\Gamma^{2,2}$. According to our
prescription, we may write the ${\bf Z}_3$-twist operator $g$ as
$$g=e^{i2\pi v^A\eta^{AB}p^B+i{\pi\over2}p^AC^{AB}p^B}g_U,\eqno(6-44)$$
where $v^A=(v^I_L,v^I_R)$, $p^A=(p^I_L,p^I_R)$ and the symmetric matrix
$C^{AB}$ is defined by
$$p^AC^{AB}{p'}^B=p^A(M-U^TMU)^{AB}{p'}^B \quad {\rm mod} \ 2,\eqno
(6-45)$$
for $p^A, {p'}^A \in \Gamma^{2,2}$. The matrices $M^{AB}$ and $U^{AB}$ are
defined in eqs.(6-11). To explicitly determine the symmetric matrix
$C^{AB}$, let us introduce a basis of $\Gamma^{2,2}$,
$$\Gamma^{2,2}=\{p^A=\sum_{a=1}^4n_ae_a^A, \ n_a\in{\bf Z} \},\eqno
(6-46)$$
where
$$e^A_i=(\mu^{iI},\mu^{iI}),\quad i=1,2,$$
$$e^A_{i+2}=(0,\alpha_i^I),\quad i=1,2.\eqno(6-47)$$
In terms of $e_a$ $(a=1,\cdots,4)$, we find
$$e^A_a(M-U^TMU)^{AB}e_b^B=\pmatrix{0 & 0 & -1 & 0 \cr
                                    0 & 0 & 0 & -1 \cr
                                    1 & 0 & 0 &  0 \cr
                                    0 & 1 & 0 &  0 \cr}_{ab}.\eqno(6-48)$$
Thus, we may choose the symmetric matrix $C^{AB}$ as
$$e^A_aC^{AB}e^B_b=\pmatrix{0 & 0 & 1 & 0 \cr
                            0 & 0 & 0 & 1 \cr
                            1 & 0 & 0 & 0 \cr
                            0 & 1 & 0 & 0 \cr}_{ab}.\eqno(6-49)$$
Since we want to construct a ${\bf Z}_3$-orbifold model, we must require
that $g^3=1$, which is equivalent to
$$\sum_{\ell=0}^2\{2v\cdot\eta U^\ell p+{1\over2}p\cdot U^{-\ell}CU^\ell p
\}=0 \quad {\rm mod} \ 2,\eqno(6-50)$$
for all $p\in\Gamma^{2,2}$. Let ${e^a}^*$ $(a=1,\cdots,4)$ be the dual
basis of $e_a$ (i.e., $e_a\cdot{e^b}^*=\delta_a^b$).
In terms of ${e^a}^*$, we may write
$$v^A=\sum_{a=1}^4y_a{e^a}^{A*}.\eqno(6-51)$$
The condition (6-50) puts a constraint on $y_a$ ($a=1,\cdots,4$) and is
equivalently written as
$$y_3={1\over3}(2\ell-\ell'),$$
$$y_4={1\over3}(-\ell+2\ell'),\quad \ell,\ell'\in {\bf Z},\eqno(6-52)$$
while $y_1$ and $y_2$ are arbitrary.

Let us consider the partition function of the untwisted sector twisted by
$g$,
$$Z(1,g;\tau)={\rm Tr}[ge^{i2\pi\tau(L_0-{2\over24})-i2\pi{\bar\tau}(\bar
L_0-{2\over24})}]_{untwist}.\eqno(6-53)$$
The zero mode part of $Z(1,g;\tau)$ is given by
$$Z(1,g;\tau)_{zero \ mode}=\sum_{k_R\in\Gamma^{0,2}_g}e^{-i2\pi v_R\cdot
k_R}e^{-i\pi{\bar\tau}k_R^2},\eqno(6-54)$$
where $\Gamma^{0,2}_g=\Lambda_R$ and $v^I_R$ can be written, in terms of
$y_a$, as
$$v^I_R=(-{1\over\sqrt2}(y_3+y_4),-{1\over\sqrt6}(y_3-y_4)).\eqno(6-55)$$
Note that the term ${1\over2}p^AC^{AB}p^B$ in eq.(6-44) does not
contribute to $Z(1,g;\tau)$ at all because
$${1\over2}p^AC^{AB}p^B=0 \quad {\rm mod } \ 2 \quad {\rm for \ all} \ p^A
\in\Gamma^{0,2}_g.\eqno(6-56)$$

According to the arguments in section 4, we can know information about
the zero mode in the $g^{-1}$-sector: The degeneracy of the ground state
is
$${\sqrt{det'(1-U)}\over V_{\Gamma^{0,2}_g}}={\sqrt3\over\sqrt3}=1,\eqno
(6-57)$$
and the momentum eigenvalues of the $g^{-1}$-sector is given by
$$(q_L,q_R)\in(0,\Lambda_W-v_R).\eqno(6-58)$$
The left-right level matching condition for $Z(g^{-1},1;\tau)$ is
$$Z(g^{-1},1;\tau+3)=Z(g^{-1},1;\tau),\eqno(6-59)$$
which is equivalent to the condition
$$3(v^I_R)^2 = {2\over3} \quad {\rm mod} \ 2.\eqno(6-60)$$
It follows from eqs.(6-52) and (6-55) that the condition (6-60) can be
rewritten as
$${2\over3}(\ell^2+{\ell'}^2-\ell\ell')={2\over3} \quad {\rm mod} \ 2.
\eqno(6-61)$$
Since the left-right level matching condition (6-59) is always a necessary
and also sufficient condition for any ${\bf Z}_3$-orbifold model, we
conclude that the one loop partition function is modular invariant if the
shift vector $(v^I_L,v^I_R)$ in eq.(6-44) satisfies (6-60) with (6-52).

In this orbifold model, to ensure modular invariance we need a nonzero
shift vector satisfying (6-60) with (6-52). It is consistent with the
argument of ref.[30]. An explicit example of the shift vector will be
given by
$$(v^I_L,v^I_R)=(0,-{1\over3}\alpha_1^I),\eqno(6-62)$$
which corresponds to $y_1=y_2=0$, $\ell=1$ and $\ell'=0$. As noted before,
the following choice of the twist operator
$$g'=e^{i2\pi v^A\eta^{AB}p^B}g_U,\eqno(6-63)$$
would also give a modular invariant partition function because of eq.
(6-56) but it does not guarantee modular invariance of correlation
functions because the twist operator $g'$ destroys the duality relation of
vertex operators.

\vskip 10mm

\section{7. Discussions}

In this paper, we have investigated the following three consistency
conditions in detail: (i) the invariance of the energy-momentum tensors
under the action of the twist operators, (ii) the duality of amplitudes
and (iii) modular invariance of partition functions. From the analysis of
the second condition (ii), we have obtained various important results. The
following two points are probably main results of this paper: The first
point is the discovery of the third term in eq.(3-25), which has to be
included as a momentum-dependent phase in the twist operator $g$ of the
untwisted sector to preserve the duality of amplitudes under the action of
$g$ and which plays an important role in modular invariance of partition
functions.
The second point is that the first condition (i) is not sufficient to
determine the allowed action of $g$ on the string coordinate and indeed
the condition (i) puts no constraint on $\Phi(p_L,p_R)$ in eq.(2-20) or
(2-23). The second condition (ii) has been found to be crucial to restrict
the allowed form of $\Phi(p_L,p_R)$ to eq.(3-25).

We have succeeded in obtaining the general class of bosonic orbifold
models. The generalization to superstring theories will be straightforward
because fermionic fields will definitely transform under the action of
twist operators.

We have restricted our considerations mainly to the untwisted sector.
However, much information about twisted sectors, in particular, zero
modes, can be obtained through modular transformations. Such information
is sufficient to obtain the partition function of the $g$-sector, $Z(g,1;
\tau)$ but not $Z(g,h;\tau)$ in general because we have not constructed
twist operators in each twisted sector. The twist operator $g$ in the
$g$-sector can, however, be found to be of the form
$$g=e^{i2\pi(L_0-\bar L_0)}.\eqno(7-1)$$
This follows from the relation
$$Z(g,g;\tau)=Z(g,1;\tau+1).\eqno(7-2)$$
To obtain an explicit operator representation of any twist operator in
every twisted sector, we may need to construct vertex operators in every
twisted sector as in the untwisted sector. In the construction of vertex
operators in twisted sectors, the most subtle part is a realization of
cocycle operators. In the case of $\xi=1$ in eq.(3-17), (untwisted state
emission) vertex operators in any twisted sector have already been
constructed with correct cocycle operators in ref.[18]. In the case of $
\xi\ne1$, the prescription given in ref.[18] will be insufficient to
obtain desired vertex operators because the duality relation will not be
satisfied. Some attempts [31] have been made but the general construction
of correct vertex operators is still an open problem.

\endpage

\section{Appendix A}

In this appendix, we present various useful formulas which will be used
in the text.

We first introduce the theta function
$$\vartheta_{ab}(\nu|\tau)=\sum_{n=-\infty}^\infty exp\{i\pi(n+a)^2\tau+i2
\pi(n+a)(\nu+b)\}.\eqno(A-1)$$
The four Jacobi theta functions are given by
$$\eqalign{
\vartheta_1(\nu|\tau)&=\vartheta_{{1\over2}{1\over2}}(\nu|\tau),\cr
\vartheta_2(\nu|\tau)&=\vartheta_{{1\over2}0}(\nu|\tau),\cr
\vartheta_3(\nu|\tau)&=\vartheta_{00}(\nu|\tau),\cr
\vartheta_4(\nu|\tau)&=\vartheta_{0{1\over2}}(\nu|\tau).\cr}\eqno(A-2)$$
They satisfy
$$\eqalign{
\vartheta_1(\nu+1|\tau)&=-\vartheta_1(\nu|\tau),\cr
\vartheta_2(\nu+1|\tau)&=-\vartheta_2(\nu|\tau),\cr
\vartheta_3(\nu+1|\tau)&=\vartheta_3(\nu|\tau),\cr
\vartheta_4(\nu+1|\tau)&=\vartheta_4(\nu|\tau),\cr}\eqno(A-3)$$
$$\eqalign{
\vartheta_1(\nu+\tau|\tau)&=-e^{-i\pi(\tau+2\nu)}\vartheta_1(\nu|\tau),\cr
\vartheta_2(\nu+\tau|\tau)&=e^{-i\pi(\tau+2\nu)}\vartheta_2(\nu|\tau),\cr
\vartheta_3(\nu+\tau|\tau)&=e^{-i\pi(\tau+2\nu)}\vartheta_3(\nu|\tau),\cr
\vartheta_4(\nu+\tau|\tau)&=-e^{-i\pi(\tau+2\nu)}\vartheta_4(\nu|\tau),
\cr}\eqno(A-4)$$
$$\eqalign{
\vartheta_1(\nu|\tau+1)&=e^{i{\pi\over4}}\vartheta_1(\nu|\tau),\cr
\vartheta_2(\nu|\tau+1)&=e^{i{\pi\over4}}\vartheta_2(\nu|\tau),\cr
\vartheta_3(\nu|\tau+1)&=\vartheta_4(\nu|\tau),\cr
\vartheta_4(\nu|\tau+1)&=\vartheta_3(\nu|\tau),\cr}\eqno(A-5)$$
$$\eqalign{
\vartheta_1(\nu/\tau|-1/\tau)&=-i(-i\tau)^{1/2}e^{i\pi\nu^2/\tau}
\vartheta_1(\nu|\tau),\cr
\vartheta_2(\nu/\tau|-1/\tau)&=(-i\tau)^{1/2}e^{i\pi\nu^2/\tau}\vartheta_4
(\nu|\tau),\cr
\vartheta_3(\nu/\tau|-1/\tau)&=(-i\tau)^{1/2}e^{i\pi\nu^2/\tau}\vartheta_3
(\nu|\tau),\cr
\vartheta_4(\nu/\tau|-1/\tau)&=(-i\tau)^{1/2}e^{i\pi\nu^2/\tau}\vartheta_2
(\nu|\tau).\cr}\eqno(A-6)$$
It is known that the Jacobi theta functions can be expanded as
$$\eqalign{
\vartheta_1(\nu|\tau)&=-2q^{1/4}f(q)sin\pi\nu\prod_{n=1}^\infty
(1-2q^{2n}cos2\pi\nu+q^{4n}),\cr
\vartheta_2(\nu|\tau)&=2q^{1/4}f(q)cos\pi\nu\prod_{n=1}^\infty(1
+2q^{2n}cos2\pi\nu+q^{4n}),\cr
\vartheta_3(\nu|\tau)&=f(q)\prod_{n=1}^\infty(1+2q^{2n-1}cos2\pi\nu
+q^{4n-2}),\cr
\vartheta_4(\nu|\tau)&=f(q)\prod_{n=1}^\infty(1-2q^{2n-1}cos2\pi\nu
+q^{4n-2}),\cr}\eqno(A-7)$$
where
$$q=e^{i\pi\tau},$$
$$f(q)=\prod_{n=1}^\infty(1-q^{2n}).\eqno(A-8)$$
Another important function is the Dedekind $\eta$-function
$$\eta(\tau)=q^{1/12}\prod_{n=1}^\infty(1-q^{2n}),\eqno(A-9)$$
which satisfies
$$\eta(\tau+1)=e^{i{\pi\over12}}\eta(\tau),$$
$$\eta(-{1\over\tau})=(-i\tau)^{1/2}\eta(\tau).\eqno(A-10)$$
We finally give the Poisson resummation formula, which will play a key
role in the modular transformation $\tau\rightarrow-{1\over\tau}$. Let $
\Gamma^{d,\bar d}$ be a $d+\bar d$-dimensional lorentzian lattice and ${
\Gamma^{d,\bar d}}^*$ be its dual lattice. Then, the formula is given by
$$\eqalign{
&\sum_{(k_L,k_R)\in\Gamma^{d,\bar d}}e^{-i{\pi\over\tau}(k_L+v_L)^2+i{\pi
\over\bar\tau}(k_R+v_R)^2}\cr
&={(-i\tau)^{d/2}(i\bar\tau)^{\bar d/2}\over V_{\Gamma^{d,\bar d}}}\sum_{
(q_L,q_R)\in{\Gamma^{d,\bar d}}^*}e^{i\pi\tau q_L^2-i\pi\bar\tau q^2_R+i2
\pi(v_L\cdot q_L-v_R\cdot q_R)},\cr}\eqno(A-11)$$
where $(v_L,v_R)$ is an arbitrary $d+\bar d$-dimensional constant vector
and $V_\Gamma$ is the unit volume of the lattice $\Gamma$.

\endpage

\section{Appendix B}

In this appendix, we shall explicitly evaluate $Z(1,g;\tau)$ and $Z
(g^{-1},1;\tau)$ and show that the constant phase $\phi$ in eq.(4-13) has
to be zero.

For our purpose, it will be sufficient to consider the case of $U_R={\bf
1}$, i.e.,
$$g_U(X^I_L,X^I_R)g_U^\dagger=(U_L^{IJ}X_L^J,X_R^I).\eqno(B-1)$$
The generalization will be straightforward. Without loss of generality, we
can assume that the orthogonal matrix $U_L$ has the following form:
$$U_L^{IJ}=\pmatrix{\delta^{ab} & 0 \cr
                    0  & V^{ij} \cr}^{IJ},\eqno(B-2)$$
where $V$ is a $d\times d$ orthogonal matrix which has no eigenvalues of
1, i.e., $det(1-V)\ne0$. Here, $I,J,\cdots$ run from 1 to $D$, $a,b,
\cdots$ from 1 to $D-d$ and $i,j,\cdots$ from 1 to $d$.

We first calculate $Z(1,g;\tau)$:
$$Z(1,g;\tau)={\rm Tr}[ge^{i2\pi\tau(L_0-{D\over24})-i2\pi\bar\tau(\bar
L_0-{D\over24})}]_{untwist},\eqno(B-3)$$
where the trace is taken over the Hilbert space of the untwisted sector.
Let $exp(i2\pi\rho_a)$ be the eigenvalues of $V$, where $0<\rho_a<1$ and
$a=1,2,\cdots,d$. Since $V$ is an orthogonal matrix, the set of
eigenvalues $\{e^{i2\pi\rho_a}\}$ is identical to the set of $\{e^{-i2\pi
\rho_a}\}$. Thus, we may write the eigenvalues of $V$ as
\footnotes{$^{*}$}{Here, we have assumed that the number of the eigenvalue
$-1$ (i.e., $\rho_a={1\over2}$) is even for simplicity.}
$$\{e^{i2\pi\rho_a} \ {\rm and } \ e^{-i2\pi\rho_a}, \ a=1,\cdots,{d
\over2}\}.\eqno(B-4)$$
The Virasoro zero mode operators $L_0$ and $\bar L_0$ in the untwisted
sector are given by
$$L_0={1\over2}(p^I_L)^2+\sum_{n=1}^\infty\alpha^I_{L-n}\alpha^I_{Ln},$$
$$\bar L_0={1\over2}(p^I_R)^2+\sum_{n=1}^\infty\alpha^I_{R-n}
\alpha^I_{Rn}.\eqno(B-5)$$
In section 3, we have seen that the twist operator $g$ would be of the
form
$$g=e^{i\pi\Phi(p)}g_U,\eqno(B-6)$$
where
$$\Phi(p)=\phi+2v^A\eta^{AB}p^B+{1\over2}p^AC^{AB}p^B.\eqno(B-7)$$
The action of $g_U$ is defined as follows:
$$g_U(X^I_L(z),X^I_R(\bar z))g_U^\dagger=(U_L^{IJ}X_L^J(z),X_R^I(\bar z)),
\eqno(B-8)$$
$$g_U|0>=|0>,\eqno(B-9)$$
where $|0>$ denotes the vacuum of the untwisted sector. The zero mode part
of $Z(1,g;\tau)$ can easily be evaluated and the result is
$$Z(1,g;\tau)_{zero \ mode}=\sum_{(k_L,k_R)\in\Gamma^{D-d,D}_g}e^{i\pi\phi
+i2\pi(v_L+v'_L)\cdot k_L-i2\pi(v_R+v'_R)\cdot k_R}e^{i\pi\tau k_L^2-i\pi
\bar\tau k_R^2},\eqno(B-10)$$
where $(v'_L,v'_R)$ is defined in eq.(4-19) and $\Gamma^{D-d,D}_g$ is the
$g$-invariant sublattice of $\Gamma^{D,D}$, i.e.,
$$\Gamma^{D-d,D}_g=\{(k_L,k_R)\in\Gamma^{D,D} \ | \ (U_Lk_L,k_R)=(k_L,k_R)
\}.\eqno(B-11)$$
Since the twist operator $g$ acts on the oscillators as
$$g(\alpha^I_{Ln},\alpha^I_{Rn})g^\dagger=(U_L^{IJ}\alpha_{Ln}^J,
\alpha_{Rn}^I),\eqno(B-12)$$
and the eigenvalues of $V$ are given by (B-4), the remaining oscillator
part of $Z(1,g;\tau)$ will be given by
$$\eqalign{&Z(1,g;\tau)_{osc}\cr
 &=|q|^{-{D\over6}}\prod_{n=1}^\infty(1-q^{2n})^{-(D-d)}(1-\bar
q^{2n})^{-D}\prod_{a=1}^{d/2}\prod_{n=1}^\infty[(1-e^{i2\pi\rho_a}q^{2n})
(1-e^{-i2\pi\rho_a}q^{2n})]^{-1},\cr}\eqno(B-13)$$
where $q=e^{i\pi\tau}$. Thus, $Z(1,g;\tau)$ can be written as
$$\eqalign{
Z(1,g;\tau)&={e^{i\pi\phi}\over|\eta(\tau)|^{2D}}\prod_{a=1}^{d/2}[{-2sin(
\pi\rho_a)(\eta(\tau))^3\over\vartheta_1(\rho_a|\tau)}]\cr
&\quad\times\sum_{(k_L,k_R)\in\Gamma^{D-d,D}_g}e^{i2\pi(v_L+v'_L)\cdot
k_L-i2\pi(v_R+v'_R)\cdot k_R}e^{i\pi\tau k_L^2-i\pi\bar\tau k_R^2},\cr}
\eqno(B-14)$$
where $\vartheta_1(\nu|\tau)$ and $\eta(\tau)$ are the Jacobi theta
function and the Dedekind $\eta$-function defined in appendix A.

Let us next consider $Z(g^{-1},1;\tau)$
$$Z(g^{-1},1;\tau)={\rm Tr}[e^{i2\pi\tau(L_0-{D\over24})-i2\pi\bar\tau(
\bar L_0-{D\over24})}]_{g^{-1}-sector},\eqno(B-15)$$
where the trace is taken over the Hilbert space of the $g^{-1}$-sector. As
discussed in section 4, the degeneracy of the ground state in the
$g^{-1}$-sector is given by
$${\sqrt{det(1-V)}\over V_{\Gamma^{D-d,D}_g}},\eqno(B-16)$$
and the eigenvalues of the momentum $(q_L,q_R)$ in the $g^{-1}$-sector are
of the form
$$(q_L,q_R)\in {\Gamma^{D-d,D}_g}^*-(v_L^*+{v_L'}^*,v_R^*+{v'_R}^*),\eqno
(B-17)$$
where
$$(v_L^*+{v_L'}^*,v_R^*+{v'_R}^*)\equiv{\cal P}_U\cdot(v_L+v_L',v_R+v'_R).
\eqno(B-18)$$
Here, ${\cal P}_U$ is the projection matrix defined in eq.(4-24). This
information about the zero mode in the $g^{-1}$-sector is sufficient to
obtain $Z(g^{-1},1,;\tau)$. The zero mode part of $Z(g^{-1},1;\tau)$ will
be given by
$$\eqalign{&Z(g^{-1},1;\tau)_{zero \ mode}\cr
  &={\sqrt{det(1-V)}\over V_{\Gamma^{D-d,D}_g}}e^{i2\pi\tau
\varepsilon_{g^{-1}}}\sum_{(q_L,q_R)\in{\Gamma^{D-d,D}_g}^*-(v_L^*
+{v_L'}^*,v_R^*+{v'_R}^*)}e^{i\pi\tau q_L^2-i\pi\bar\tau q_R^2},\cr}\eqno
(B-19)$$
where $\varepsilon_{g^{-1}}$ denotes the zero point energy (of the left
mover) of the ground state in the $g^{-1}$-sector,
$$\varepsilon_{g^{-1}}=2\sum_{a=1}^{d/2}{1\over4}\rho_a(1-\rho_a).\eqno
(B-20)$$
In the $g^{-1}$-sector, $d$ of $D$ oscillators of the left mover are
twisted with the phases (B-4). Thus, the remaining oscillator part of $Z
(g^{-1},1;\tau)$ will be given by
$$\eqalign{&Z(g^{-1},1;\tau)_{osc}\cr
&=|q|^{-{D\over6}}\prod_{n=1}^\infty(1-q^{2n})^{-(D-d)}(1-\bar
q^{2n})^{-D}\prod_{a=1}^{d/2}\prod_{n=1}^\infty[(1-q^{2(n-\rho_a)})(1-q^{2
(n-1+\rho_a)})]^{-1}.\cr}\eqno(B-21)$$
Therefore, $Z(g^{-1},1;\tau)$ can be written as
$$\eqalign{
Z(g^{-1},1;\tau)&={\sqrt{det(1-V)}\over V_{\Gamma^{D-d,D}_g}}{e^{i2\pi\tau
\varepsilon_{g^{-1}}}\over|\eta(\tau)|^{2D}}\prod_{a=1}^{d/2}[{-ie^{-i\pi
\tau\rho_a}(\eta(\tau))^3\over\vartheta_1(\rho_a\tau|\tau)}]\cr
&\quad\times \sum_{(q_L,q_R)\in{\Gamma^{D-d,D}_g}^*-(v_L^*+{v_L'}^*,v_R^*
+{v'_R}^*)}e^{i\pi\tau q_L^2-i\pi\bar\tau q_R^2}.\cr}\eqno(B-22)$$

{}From the expressions (B-14) and (B-22), it is easy to see that
$$Z(1,g;-{1\over\tau})|_{\phi=0}=Z(g^{-1},1;\tau).\eqno(B-23)$$
This proves that the phase $\phi$ of the twist operator $g$ in the
untwisted sector has to vanish. To show eq.(B-23), we may use the formulas
in appendix A. We can easily find
$$\eqalign{
Z(1,g,1;-{1\over\tau})&={e^{i\pi\phi}\over|\eta(\tau)|^{2D}}\prod_{a
=1}^{d/2}[{-2isin(\pi\rho_a)(\eta(\tau))^3\over e^{i\pi\tau(\rho_a)^2}
\vartheta_1(\rho_a\tau|\tau)}]\cr
&\quad\times {1\over V_{\Gamma^{D-d,D}_g}}\sum_{(q_L,q_R)\in{
\Gamma^{D-d,D}_g}^*-(v_L^*+{v_L'}^*,v_R^*+{v'_R}^*)}e^{i\pi\tau q_L^2-i\pi
\bar\tau q_R^2},\cr}\eqno(B-24)$$
where we have used the relation
$$(v_L+v'_L)\cdot k_L-(v_R+v'_R)\cdot k_R=(v_L^*+{v'_L}^*)\cdot k_L-(v_R^*
+{v'_R}^*)\cdot k_R,\eqno(B-25)$$
for $(k_L,k_R)\in\Gamma^{D-d,D}_g$. Using eq.(B-20) and the relation
$$\sqrt{det(1-V)}=\prod_{a=1}^{d/2}(2sin(\pi\rho_a)),\eqno(B-26)$$
we finally obtain the relation (B-23).

\endpage

\references

\item{[1]} L. Dixon, J.A. Harvey, C. Vafa and E. Witten, Nucl. Phys. {\bf
B261} (1985) 678; {\bf B274} (1986) 285.
\item{[2]} A. Font, L.E. Ib\'a\~ nez, F. Quevedo and A. Sierra, Nucl.
Phys. {\bf B331} (1990) 421.
\item{[3]} J.A. Casas and C. Mu\~ noz, Nucl. Phys. {\bf B332} (1990) 189.
\item{[4]} Y. Katsuki, Y. Kawamura, T. Kobayashi, N. Ohtsubo, Y. Ono and
K. Tanioka, Nucl. Phys. {\bf B341} (1990) 611.
\item{[5]} A. Fujitsu, T. Kitazoe, M. Tabuse and H. Nishimura, Intern. J.
Mod. Phys. {\bf A5} (1990) 1529.
\item{[6]} H. Kawai, D. Lewellyn and A.H. Tye, Phys. Rev. Lett. {\bf 57}
(1986) 1832; Nucl. Phys. {\bf B288} (1987) 1;
\item{   } I. Antoniadis, C. Bachas and C. Kounnas, Nucl. Phys. {\bf B289}
(1987) 87.
\item{[7]} W. Lerche, A.N. Schellenkens and N.P. Warner, Phys. Rep. {\bf
177} (1989) 1.
\item{[8]} D. Gepner, Phys. Lett. {\bf B199} (1987) 380; Nucl. Phys. {\bf
B296} (1987) 757.
\item{[9]} Y. Kazama and H. Suzuki, Nucl. Phys. {\bf B321} (1989) 232.
\item{[10]} C. Vafa and N.P. Warner, Phys. Lett. {\bf B218} (1989) 51;
\item{    } W. Lerche, C. Vafa and N.P. Warner, Nucl. Phys. {\bf B324}
(1989) 427;
\item{    } P.S. Howe and P.C. West, Phys. Lett. {\bf B223} (1989) 377; {
\bf B244} (1989) 270.
\item{[11]} E.S. Fradkin and A.A. Tseytlin, Phys. Lett. {\bf B158} (1985)
316; Nucl. Phys. {\bf B261} (1985) 1;
\item{    } C.G. Callan, D. Friedan, E.J. Martinec and M.J. Perry, Nucl.
Phys. {\bf B262} (1985) 593;
\item{    } C.G. Callan, I.R. Klebanov and M.J. Perry, Nucl. Phys. {\bf
B278} (1986) 78;
\item{    } T. Banks, D. Nemeschansky and A. Sen, Nucl. Phys. {\bf B277}
(1986) 67.
\item{[12]} P. Candelas, G. Horowitz, A. Strominger and E. Witten, Nucl.
Phys. {\bf B258} (1985) 46.
\item{[13]} D.J. Gross and V. Periwal, Phys. Rev. Lett. {\bf 60} (1988)
2105.
\item{[14]} M. Kaku and K. Kikkawa, Phys. Rev. {\bf D10} (1974) 1110,1823;
\item{    } H. Hata, K. Itoh, T. Kugo, H. Kunitomo and K. Ogawa, Phys.
Rev. {\bf D35} (1987) 1318,1356;
\item{    } E. Witten, Nucl. Phys. {\bf B268} (1986) 253.
\item{[15]} M. Sakamoto and M. Tabuse, Kobe preprint, KOBE-92-02 (1992).
\item{[16]} K.S. Narain, M.H. Sarmadi and C. Vafa, Nucl. Phys. {\bf B288}
(1987) 551; {\bf B356} (1991) 163.
\item{[17]} K.S. Narain, Phys. Lett. {\bf B169} (1986) 41;
\item{    } K.S. Narain, M.H. Sarmadi and E. Witten, Nucl. Phys. {\bf
B279} (1987) 369.
\item{[18]} K. Itoh, M. Kato, H. Kunitomo and M. Sakamoto, Nucl. Phys. {
\bf B306} (1988) 362.
\item{[19]} J.H. Schwarz, Phys. Rep. {\bf 8C} (1973) 269;{\bf 89} (1982)
223;
\item{    } J. Scherk, Rev. Mod. Phys. {\bf 47} (1975) 123.
\item{[20]} D.J. Gross, J.A. Harvey, E. Martinec and R. Rohm, Nucl. Phys.
{\bf B256} (1985) 253; {\bf B267} (1986) 75.
\item{[21]} J.L Cardy, Nucl. Phys. {\bf B270} (1986) 186.
\item{[22]} H. Suzuki and A. Sugamoto, Phys. Rev. Lett. {\bf 57} (1986)
1665.
\item{[23]} N. Sakai and Y. Tanii, Nucl. Phys. {\bf B287} (1987) 457.
\item{[24]} K. Inoue, S. Nima and H. Takano, Prog. Theor. Phys. {\bf 80}
(1988) 881.
\item{[25]} I. Frenkel and V. Ka\v c, Invent. Math. {\bf 62} (1980) 23;
\item{    } G. Segal, Commun. Math. Phys. {\bf 80} (1981) 301;
\item{    } P. Goddard and D. Olive, Intern. J. Mod. Phys. {\bf A1} (1986)
303.
\item{[26]} V.A. Kosteleck\'y, O. Lechtenfeld, W. Lerche, S. Samuel and S.
Watamura, Nucl. Phys. {\bf B288} (1987) 173.
\item{[27]} M. Sakamoto, Phys. Lett. {\bf B231} (1989) 258.
\item{[28]} C. Vafa, Nucl. Phys. {\bf B273} (1986) 592.
\item{[29]} K. Inoue, M. Sakamoto and H. Takano, Prog. Theor. Phys. {\bf
78} (1987) 908.
\item{[30]} Y. Imamura, M. Sakamoto and M. Tabuse, Phys. Lett. {\bf B266}
(1991) 307.
\item{[31]} J. Erler, D. Jungnickel, J. Lauer and J. Mas, preprint
SLAC-PUB-5602.

\end